# Numerical Investigation of Chemically Reacting Rarefied Hypersonic Flows Over A Backward-Facing Step

Thesis
Submitted in partial fulfillment of the requirements of
BITS F421T Thesis

By

**Arjun Singh**

**ID No. 2017A4TS0707H**

Under the supervision of

**Dr K. Ram Chandra Murthy**
**Assistant Professor**

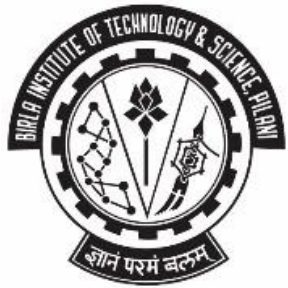

**BIRLA INSTITUTE OF TECHNOLOGY AND SCIENCE, PILANI
HYDERABAD CAMPUS**

**(December 2020)**

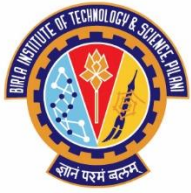

<div style="text-align:right">(Date)</div>

# CERTIFICATE

This is to certify that the thesis entitled "Numerical Investigation of Chemically Reacting Rarefied Hypersonic Flows Over A Backward-Facing Step" submitted by <u>Arjun Singh</u> ID No 2017A4TS0707H in partial fulfilment of the requirement of BITS F421T Thesis embodies the original work done by him under my supervision.

**Dr K. Ram Chandra Murthy**
**Assistant Professor**
Date:

i

# Acknowledgments

This page goes out to all those without whom this work would not have been possible, and it is therefore impossible to begin without thanking Dr K. Ram Chandra Murthy, my mentor for this work. He has been patient, interactive, and has always given whatever was required, every single time. He is the source of strength and inspiration for this work, and any amount of credit given to him would not be enough. I thank him for this opportunity to learn what research is all about, and how to work to produce tangible results. I would also like to thank Mr Deepak Nabapure, who has provided the results for non-reacting flow.

I would also like to gratefully acknowledge the High-Performance Computing (HPC) Laboratory, BITS Pilani, Hyderabad Campus, for their technical assistance and support in providing the computing facilities to perform the numerical simulations

I am also grateful to my parents Ravinder Singh and Kavita Sambyal, and my friend Richa, for their constant support in many forms.

# Abstract

The present study employs the Direct Simulation Monte Carlo (DSMC) method, a widely used numerical method for studying non-continuum flows, to investigate the flow characteristics of a chemically reacting hypersonic flow over a backward-facing step. This work explores the effects of the Knudsen number on the behaviour of rarefied flow and compares the results with the non-reacting flow. The primary objective of the present thesis is to elucidate the variation of temperature in the flow domain with Knudsen number. The results of the study show that chemical reactions affect the temperature distribution in the flow domain significantly.

# List of Figures

# List of Tables

# Nomenclature

**Symbols**

| | |
|---|---|
| $A$ | : Surface area [m$^2$] |
| $c$ | : Molecular velocity [m/s] |
| $C_p$ | : Pressure coefficient |
| $C_f$ | : Skin friction coefficient |
| $C_h$ | : Heat transfer coefficient |
| $d$ | : Particle diameter [m] |
| $f$ | : Velocity distribution function |
| $F_N$ | : Number of real molecules represented by each simulated molecule |
| $h$ | : Characteristic length- (Step height) [m] |
| H | : Channel height at outlet [m] |
| $Kn$ | : Knudsen Number |
| $m$ | : Molecular mass [kg] |
| $Ma$ | : Mach Number |
| $n$ | : Number density [m$^{-3}$] |
| *N* | : Number of molecules |
| *p* | : Pressure [N/m$^2$] |
| *Re* | : Reynolds Number |
| *t* | : Time [s] |
| *T* | : Temperature [K] |
| X | : Axial distance in *x*-direction [m] |
| Y | : Axial distance in *y*-direction [m] |

## Greek Symbols

| | |
|---|---|
| $\chi$ | : Mole fraction |
| $\sigma$ | : Collision cross section [m$^2$] |
| $\Omega$ | : Solid angle [sr] |
| $\omega$ | : Viscosity-temperature index |
| $\rho$ | : Density of the gas [kg/m$^3$] |
| $\mu$ | : Viscosity of the gas [N.s/m$^2$] |
| $\tau$ | : Shear stress [N/m$^2$] |
| $\lambda$ | : Mean free path [m] |
| $\zeta$ | : Degree of freedom |
| $\delta$ | : Boundary layer thickness |

## Subscripts

| | |
|---|---|
| $\infty$ | : Free-stream value |
| *w* | : Wall value |
| *r* | : Relative value |
| * | : Post-collision value |
| 0 | : Overall value |
| $inc$ | : Incident molecule |
| $ref$ | : Reflected molecule |

## Abbreviations

| | |
|---|---|
| BE | : Boltzmann Equation |
| DSMC | : Direct Simulation Monte Carlo |
| BFS | : Backward-Facing Step |
| MD | : Molecular Dynamics |
| PPC | : Particles Per Cell |
| Q-K | : Quantum-Kinetic |

# Contents

# Chapter 1

# Introduction

## 1.1 Background

Re-entry vehicles entering the earth's atmosphere operate at high altitudes reaching hypersonic velocities. Atmospheric altitudes between 20 kilometres to 100 kilometres are generally regarded as near-space. At such high altitudes, a rarefied environment prevails. Typically, these hypersonic vehicles encounter extreme temperatures. As a result, the vibrational modes of the ambient gas species are excited, and chemical reactions occur. Therefore, a detailed understanding of the aerothermodynamics of such flows is crucial for the design of the re-entry vehicle's thermal protection system (TPS). During the design stage or in motion, these TPS are prone to surface irregularities which can be simplified to geometries such as cavities, steps, and gaps. A backward-facing step (BFS) is one such irregularity that leads to flow separation and significantly alters the heat flux distribution on aerospace vehicles' surface.

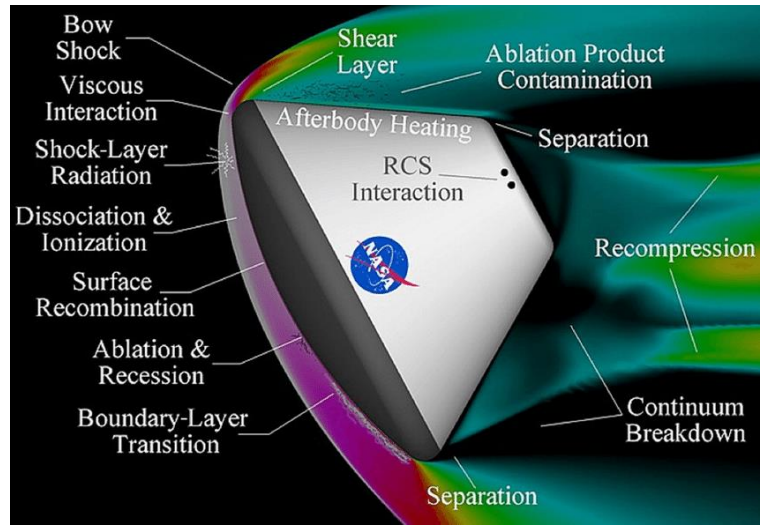

**Fig. 1    Schematic view of a typical planetary re-entry phenomena**

## 1.2 Literature Survey

The BFS configuration has been widely studied due to interesting flow behaviours such as flow separation and reattachment of flow over the step. Over the years studies have employed numerical and analytical methods for 3D and 2D flows to investigate the effect of aspect ratio [1], Prandtl Number [2], Reynold's Number [3], and step height [4], and have helped gain more insight on the flow physics in a BFS configuration. Flow over BFS has also been widely studied in different rarefaction regimes.

*1.2.1   BFS flow in Slip Regime*

In the slip regime, Choi et al. [5] implemented a Langmuir slip boundary condition combined with the continuum-based compressible Navier-Stokes equation for the flow over BFS and accurately predicted the flow physics. Hsieh [6] investigated the 3D microscale BFS flow using the DSMC method by simplifying 3D flows to 2D flow by maintaining the aspect ratio greater than 5. The study showed that for an inlet Kn of 0.04 and an aspect ratio of less than 1, the flow separation, recirculation, and reattachment disappeared. Beskok [7] developed a velocity-slip boundary condition based on obtaining the information of slip at a distance of one mean free path away from the surface and applied the same for the gas flows over BFS. The results, when compared the same with the DSMC findings, were found them to agree with each other. Celik and Edis [8] formulated a characteristic based split Navier Stokes FEM solver and studied the rarefied flow over a through a BFS duct. Baysal et al. [9] also used the Navier-Stokes equation with slip/jump boundary conditions and investigated the control of separated flow past a BFS. Xue and Chen [10] studied the effect of Kn on micro BFS using the DSMC method, both in the slip and transition regime. Their results indicated that the flow phenomena such as flow separation, reattachment, and recirculation disappeared when Kn exceeded 0.1. The reasons for the above phenomena were explained by Xue et al. [11] in another study.

*1.2.2   BFS flow in the Transition Regime*

In the transition regime, Kursun and Kapat [12] numerically studied the rarefied flow through BFS using the DSMC-IP (Information preservation) method. They carried out the studies for Re of 0.03 to 0.64, Ma of 0.013 to 0.083. For the above conditions, they did not report any recirculation region. Bao and Lin [13] used the continuum based Burnett equations to study the microscale-BFS by comparing their numerical results with the available experimental and numerical results from the literature and found good agreement among them. Darbandi and Roohi [14] used the DSMC method to study the subsonic flow through micro/nano-BFS for different Kn. Their results indicated a decrease in the separation region length as flow rarefaction increases as it enters the transition regime. Gavasane et al. [15] studied the Argon flow through a micro-BFS for different Kn and pressure ratios using DSMC. From the above discussion, it can be observed that most of the studies either considered monoatomic Argon or diatomic Nitrogen as the fluid in their investigations.

Based on the above survey, it is evident that there is considerable literature on the numerical study of rarefied flow over a backward-facing step in various rarefaction regimes, but mostly at the microscale. However, the aerothermodynamic surroundings that re-entry vehicles experience and the physical phenomena of such highly rarefied hypersonic flows are poorly understood.

### 1.2.3  Re-entry hypersonic rarefied flow over a BFS

Over the years, many studies have been carried out to study the hypersonic rarefied flow in the re-entry environment for different rarefaction regimes. Guo et al. [16] carried out a detailed investigation of non-reacting rarefied hypersonic flow-field characteristics over a BFS under active flow control at different atmospheric altitudes using the Direct Simulation Monte Carlo (DSMC) method. Leite et al. [17] studied the non-reacting hypersonic flow over a BFS in the transition regime using the DSMC method. Nabapure et al. [18] studied both subsonic and hypersonic non-reacting flow over a BFS in transitional and free molecular regime using the DSMC method. In another study, Nabapure et al. [19] investigated the effect of Mach number and BFS wall temperature on the flow field properties.

Although rarefied flow over a BFS in hypersonic re-entry surroundings has been explored in the past, the studies conducted have mainly centered around investigating the effects of rarefaction in the slip regime or at the onset of the transition regime for non-reacting flow. The present study explores the effect of rarefaction and chemical reactions in a hypersonic rarefied flow over a BFS which has not been explored in the past. The present study gives a detailed account of the effects of chemical reactions on variation in temperature distribution for both chemically reacting flow and non-reacting flow and compares the results between them.

## 1.3 Thesis Overview

Chapter 2 addresses the governing equations describing the behaviour of rarefied gases. The chapter then further gives a brief account of the DSMC method, the computational methodology adopted in the present study to simulate the hypersonic rarefied gas flows. Moreover, a brief description of the implemented chemical reactions along with the Q-K methodology has been provided.

Chapter 3 introduces the problem statement and discusses the freestream properties along with the computational parameters considered in the present study.

Chapter 4 aims to establish the validity of the solver used in the present study. The solver is validated by comparing with relevant data in the literature available.

Chapter 5 gives a detailed account of the results of the problem statement defined in Chapter 3.

Chapter 6 reports the key-findings and provides a brief account of studies which can be explored in the future.

# Chapter 2

# Computational Methodology

## 2.1 Governing Equations

The rarefaction of a flow can be characterized by Knudsen number ($Kn$), which is given by [20],

$$Kn = \frac{\lambda}{h} \tag{1}$$

where λ denotes the mean free path, and h represents the characteristic dimension of the geometry under consideration. The Kn is used to classify the flow into Continuum regime ($Kn \leq 0.001$), Slip regime ($0.001 \leq Kn \leq 0.1$), Transition regime ($0.1 \leq Kn \leq 10$), and Free-molecular regime ($Kn \geq 10$). In the slip flow regime, the continuum hypothesis breaks down, and the Navier

Stokes equations do not describe the flow accurately. Such flows are more accurately described by the Boltzmann equation given [21] by:

$$\frac{\partial f}{\partial t} + v \cdot \nabla_x f = \frac{1}{Kn} Q(f,f), \quad x \epsilon \mathbb{R}^{dx}, v \epsilon \mathbb{R}^{dv} \tag{2}$$

where $f(t,x,v)$ denotes the density distribution function of a dilute gas at a position $x$, velocity $v$, and at time $t$. $Kn$ represents the Knudsen number of the flow and the collision operator $Q(f,f)$ represents the binary collisions. The $Kn$, $Re$, and $Ma$ are also related to each other as follows:

$$Kn = \sqrt{\frac{\pi \gamma}{2}} \frac{Ma}{Re}, \text{ where } Ma = \frac{U_\infty}{\sqrt{\gamma RT}} \text{ and } Re = \frac{\rho U_\infty h}{\mu} \tag{3}$$

where ρ is the density, $U_\infty$ is the free-stream velocity, h is the step height, and $\mu$ is the dynamic viscosity of the fluid. The Direct Simulation Monte Carlo (DSMC) method is a powerful probabilistic technique developed by Bird [22] for studying the flow in rarefied regimes. *dsmcFoam+* [23], developed in the OpenFOAM framework, is an open-source DSMC solver and has been employed in the present study.

## 2.2 The Direct Simulation Monte Carlo (DSMC) Method

Pioneered by Bird in the early 1960s, the DSMC method is one of the most popular and successful techniques to model Rarefied Gas Flows. DSMC method utilizes probabilistic (Monte Carlo) simulation to solve the Boltzmann Equation for a finite Knudsen number and it has been developed subsequently. The DSMC currently maintains its significant role in addressing rarefied gas dynamics issues amid the emergence of alternative methods. It is used in solving diverse problems involving high-altitude aerodynamics, laser ablation, vacuum techniques, rarefied plasma, etc.

The DSMC solver also models a bulk of particles as one DSMC particle having a position, velocity, and internal energy, using the concept of N-equivalent particles. The concept of this process is to independently calculate at every timestep the movement and intermolecular collision of a finite number of fictitious particles separately, considering that each particle constitutes a considerable number of actual particles.

The physical space under investigation is represented by a computational grid, divided into several cells. Typically, the cell dimensions should be kept smaller than one-third of the local mean free path [24,25]. Since the DSMC method uses small cells within a domain, a timestep is selected such that the particle does not cross the whole width of the cell within a single timestep. Moreover, for simulating collisions between particles, the timestep must be less than the mean collision time [26,27]. For accurate results, the timestep is taken as the time taken by a DSMC particle at Vrms to cross one-fifth of a cell. Additionally, a condition of a minimum number of DSMC particles per cell should be satisfied. This is because the collision rate is directly proportional to the number of DSMC particles per cell [28]. Thus, each cell should have the largest possible number of particles. However, this brings its computational challenges. Thus, it is necessary to determine the optimum number of particles in each cell so that statistical accuracy along with a realistic computational expenditure can be maintained. For the present study, the particles per cell (PPC) were kept greater than 30.

The DSMC algorithm involves four different steps: 1. Read the grid data and define the initial and boundary conditions; 2. Calculate the number of DSMC molecules and initialize them in the domain; 3. Model the interaction of DSMC molecules with the boundaries; 4. Index the simulated DSMC molecules; 5. Use the probabilistic sampling to model the collision of the simulated DSMC molecules by; 6. Sample the flow field and repeat the steps 3-6; 7. Output the sampled flow field variables.

The initialisation of the computational domain is based on the freestream conditions of the flow. During each timestep, various physical properties like density, velocity, internal energy, and temperature are specified according to the given boundary conditions for the particles entering the computational domain.

A linear combination of the thermal velocity and the freestream velocity to calculate the velocity of the simulated particle. Initially, at time zero, the boundary conditions corresponding to the flow are imposed. After a sufficiently large time, a steady-state flow is established. Thus, a time average is calculated after reaching a steady-state to obtain the desired steady result.

The next step involves the movement of the DSMC particles in the computational domain appropriate to their velocity components and timestep size. Their new locations are thus determined. Particles which leave the computational domain are removed from the flow and thus new particles are introduced into the computational domain in their place. For surface collisions with surfaces, the treatment requires the application of the conservation laws and the application of Maxwellian velocity distribution function. These collisions can be treated as specular, diffuse or a combination of these two types. Following this, the particles are indexed by cell location so that the calculation of intermolecular collisions and sampling of the flow field can be done.

Following the deterministic treatment of their motion, the intermolecular collisions are determined. Many schemes such as Time counter (TC), Nearest neighbor (NN), Null collision,

Modified Nanbu, Ballot box, No Time Counter (NTC) [29], Bernoulli's Trials, Simplified and Generalized Bernoulli Trials, are available for the selection of collision pair. A review of these models has been elucidated in [30]. In the present study, the collision sampling pairs are determined using the No Time Counter (NTC) scheme.

The maximum number of particle pairs ($N_c$) considered for binary collisions $i$, in the NTC, is given by,

$$N_c = \frac{1}{2}\frac{N(N-1)F_N(\sigma_T c_r)_{max}\Delta t}{V_c} \qquad (4)$$

where, $F_N$ denotes the N-equivalent particles, $N$ represents the total number of particles in the cell, $(\sigma_T c_r)_{max}$ is the maximum value of the product of collision cross-section and relative particle velocity, $\Delta t$ is the timestep and $V_c$ is the volume of the cell.

Once collision pairs are selected, collisions are modelled through techniques such as Hard-Sphere (HS) model, Generalized Hard-Sphere (GHS) model, Generalized Soft-Sphere (GSS) model, Variable Hard-Sphere (VHS) model, Variable Soft-Sphere (VSS) model and Variable Sphere (VS) molecular model. A detailed description has been elucidated in [31]. The current study employs the VHS collision model to handle the intermolecular collisions. The VHS collision model describes the mean-free path ($\lambda$) of the gas as,

$$\lambda = \frac{2}{15}(5-2\omega)(7-2\omega)\left(\frac{\mu}{\rho}\right)\sqrt{\frac{m}{2\pi kT}} \qquad (5)$$

where, $m, \omega, T, \rho, \mu, k$, are the gas mass, viscosity-temperature index, gas temperature, gas density, gas viscosity, and Boltzmann constant, respectively.

Described by the VHS model, these collisions are either elastic or inelastic. During an elastic collision, the exchange of momentum and energy does take place among the particles. However, during an inelastic collision, a transfer of momentum and energy between the translational and internal modes is observed. Moreover, during the collision, for redistribution of energy between translational and internal modes, the Larsen-Borgnakke scheme is employed.

A collision number, $Z$, defined as the mean of molecular collisions for an energy mode to reach equilibrium, describes the equilibration rate of an internal energy mode with the translational mode. In the current solver, the vibrational energy of a particle can assume equally spaced discrete quantum levels in accordance with the harmonic oscillator model. A serial application of the quantum Larsen-Borngakke model is adopted to employ relaxation of vibrational modes. In this procedure, a quantized collision temperature determines the vibrational collision number. The electronic mode is out of the scope of the current work in consideration and therefore has not been employed. Finally, the standard Larsen-Borngakke procedure is employed for the rotational energy mode, and the rotational collision number is chosen as a constant value of 5.

## 2.3 Chemical Reactions

Several chemical reaction models like the total collision energy (TCE) model, introduced by Bird [32], are available in the DSMC. Particularly the popular TCE utilizes the equilibrium kinetic theory to correlate the macroscopic and microscopic properties of the gas by translating the Arrhenius rate coefficients into collision probabilities which are a function of macroscopic gas temperature and microscopic collision energy, respectively. In recent years, an alternative reaction model, the quantum-kinetic (Q-K) chemistry model was proposed by Bird [33], which has been employed in *dsmcFoam+*.

The Q-K method is a molecular level chemistry model which employs the fundamental molecular properties, antithetical to the classical TCE model which relies on experimental data availability. The Quantum-Kinetic (Q-K) chemistry model has been used to define the chemical reactions of a 5-species air model with a total of 19 chemical reactions in the *dsmcFoam+* as shown in Table 1. Relevant to the present discussion are two chemical reaction types – Dissociation (No. 1-15 in Table 1) and Exchange reactions (No. 16-19 in Table 1). The present study investigates the effects of chemical reactions on the temperature of the flow for five different Knudsen numbers: 0.05, 0.10, 1.06, 10.33, and 21.10, covering all the rarefaction regimes.

**Table 1  List of chemical reactions list employed**

| No. | Reaction | Heat of formation $\times 10^{19}$ J |
|---|---|---|
| 1. | $O_2 + N \rightarrow O + O + N$ | 8.197 |
| 2. | $O_2 + NO \rightarrow O + O + NO$ | 8.197 |
| 3. | $O_2 + N2 \rightarrow O + O + N_2$ | 8.197 |
| 4. | $O_2 + O_2 \rightarrow O + O + O_2$ | 8.197 |
| 5. | $O_2 + O \rightarrow O + O + O$ | 8.197 |
| 6. | $N_2 + O \rightarrow N + N + O$ | 15.67 |
| 7. | $N_2 + O_2 \rightarrow N + N + O_2$ | 15.67 |

| 8.  | $N_2 + NO \rightarrow N + N + NO$ | 15.67  |
|-----|-----------------------------------|--------|
| 9.  | $N_2 + N_2 \rightarrow N + N + N_2$ | 15.67  |
| 10. | $N_2 + N \rightarrow N + N + N$   | 15.67  |
| 11. | $NO + N_2 \rightarrow N + O + N_2$ | 10.43  |
| 12. | $NO + O_2 \rightarrow N + O + O_2$ | 10.43  |
| 13. | $NO + NO \rightarrow N + O + NO$  | 10.43  |
| 14. | $NO + O \rightarrow N + O + O$    | 10.43  |
| 15. | $NO + N \rightarrow N + O + N$    | 10.43  |
| 16. | $NO + O \rightarrow O_2 + N$      | 2.719  |
| 17. | $N_2 + O \rightarrow NO + N$      | 5.175  |
| 18. | $O_2 + N \rightarrow NO + O$      | -2.719 |
| 19. | $NO + N \rightarrow N_2 + O$      | -5.175 |

# Chapter 3

# Problem Definition

## 3.1 Overview

The present work explores chemically reacting rarefied hypersonic flow over a 2D BFS, for a range of Knudsen numbers. The geometry of the BFS is shown in Fig. 2. The step height (h) was fixed to 3 mm for all the cases. The inlet is located at 45 mm from the step, and the outlet at 105 mm downstream

**Table 2  Geometric parameters of the 2D-BFS**

| Parameter | Upstream length | Downstream length | Channel Height (H) | Step Height (h) | Step position |
|---|---|---|---|---|---|
| Value | 60 mm | 105 mm | 60 mm | 3 mm | X = 60 mm |

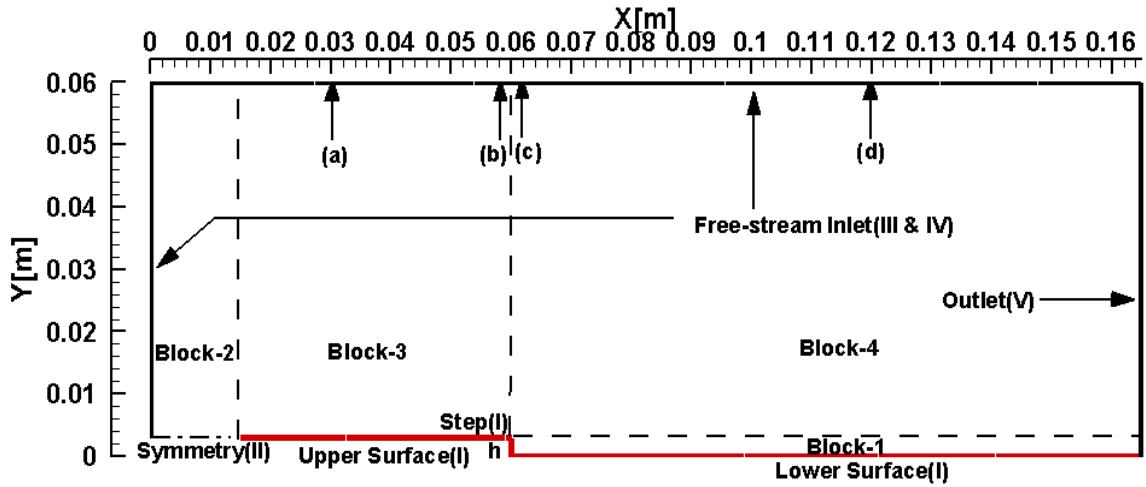

**Fig. 2　Schematic of the 2D BFS.**

## 3.2 Boundary conditions and Mesh

Table 3 shows the boundary conditions defined for all the surfaces. Side-I includes both upper and lower surfaces as well as the step. Surfaces in Side-I were defined with a diffuse reflection wall boundary condition with full thermal accommodation. A specular reflection wall boundary condition was applied to the symmetry surface placed 15h upstream of the step. Sides-III and IV represent the free-stream inlet boundaries where the molecules can freely enter and exit. Side-V represents the outlet where the molecules exit.

**Table 3　Boundary conditions**

| Surface | I | II | III | IV | V |
|---|---|---|---|---|---|
| Boundary condition | Wall | Symmetry | Free-stream inlet | Free-stream inlet | Outlet |

　　For simulating collisions, the flow domain was divided into four blocks (Blocks-1, 2, 3, 4 shown in Fig. 2) which are meshed using a structured grid comprising of quadrilateral cells (shown in Fig. 3). The dimensions of the cell $\Delta x, \Delta y$ were maintained less than $\lambda/3$ in the flow domain

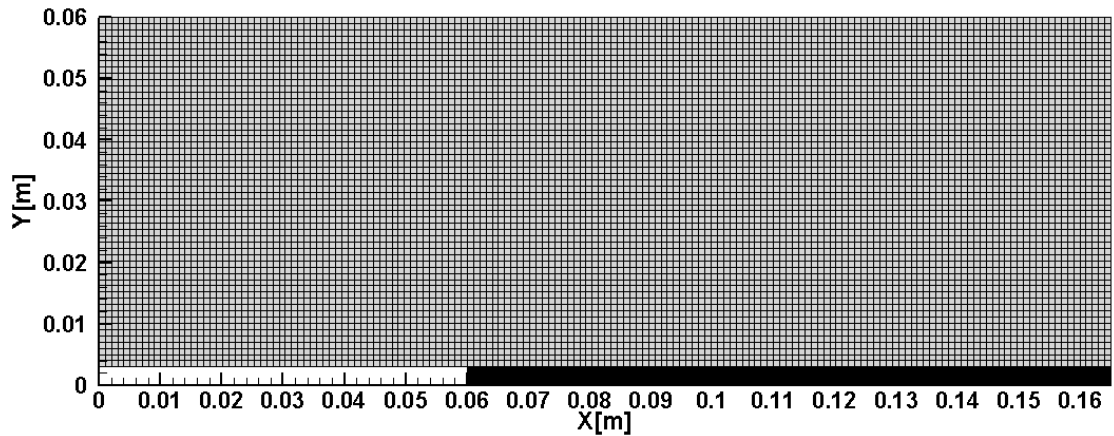

**Fig. 3　　Meshed domain.**

## 3.3 Computational Parameters

　　Table 4 and Table 5 respectively show the gas properties and freestream conditions used in the present study. The freestream flow of air consists of 76% $N_2$ and 23% $O_2$. The average molecular weight of air is approximately 28.96 g/mol. $\chi, m, d$ in Table 4 denote the mole fraction, molecular mass, molecular diameter respectively. All the cases were sampled for 75,000 timesteps with roughly 30 particles per cell and simulated till $1.5 \times 10^{-3}$ $s$ of the physical flow time. Simulations were carried out for non-reacting and chemically reacting rarefied hypersonic perfect-gas flow.

**Table 4 Gas properties [34]**

|  | $\chi$ | $m$, kg | $d$, m | $\omega$ |
|---|---|---|---|---|
| $O_2$ | 0.237 | $5.312 \times 10^{-26}$ | $4.07 \times 10^{-10}$ | 0.77 |
| $N_2$ | 0.763 | $4.650 \times 10^{-26}$ | $4.17 \times 10^{-10}$ | 0.74 |

**Table 5 Flow and simulation parameters [35]**

| Altitude | 55.02Km | 60.5Km | 77Km | 91.5Km | 95Km |
|---|---|---|---|---|---|
| Free stream velocity ($U_\infty$) [m/s] | 8050.75 | 7829.25 | 7134.5 | 6851.38 | 6887.03 |
| Free stream temperature, ($T_\infty$)[ K] | 259.39 | 245.64 | 204.49 | 186.89 | 188.84 |
| Wall temperature, ($T_w$)[ K] | 1037.56 | 982.56 | 817.96 | 747.56 | 755.36 |
| Free stream Pressure, ($p_\infty$)[N/m2] | 39.86 | 20.51 | 1.72 | 0.14 | 0.069 |

| Density, $(\rho)[kg/m^3]$ | $5.35 \times 10^{-4}$ | $2.90 \times 10^{-4}$ | $2.94 \times 10^{-5}$ | $2.61 \times 10^{-6}$ | $1.27 \times 10^{-6}$ |
|---|---|---|---|---|---|
| Number density, $(n_\infty)[\text{m-3}]$ | $1.11 \times 10^{22}$ | $5.62 \times 10^{21}$ | $5.29 \times 10^{20}$ | $5.45 \times 10^{19}$ | $2.66 \times 10^{19}$ |
| Mean free path, $(\lambda)(\text{m})$ | $1.51 \times 10^{-4}$ | $3 \times 10^{-4}$ | $3.19 \times 10^{-3}$ | $3.1 \times 10^{-2}$ | $6.33 \times 10^{-2}$ |
| Knudsen Number (Kn) | 0.05 | 0.10 | 1.06 | 10.33 | 21.10 |

# Chapter 4

# Verification and Validation

To establish the accuracy of results in the present study, the solver was validated with results in the existing literature. Firstly, validation was performed for a comparable case of non-reacting flow over a BFS, then a benchmark case of chemically reacting hypersonic flow over a circular cylinder was validated.

## 4.1 Non-reacting flow over a BFS

Guo et al. [16] studied flow characteristics of non-reacting hypersonic flow over a BFS using DSMC method. Table 6 shows the details of the free-stream conditions used in their study for an atmospheric altitude of H = 30 km. Results obtained by them were compared with the present study in Fig. 4 and Fig. 5. The Mach number contours observed by them are shown in Fig. 4(a) and show a good visual agreement with the Mach number contours obtained in the present study, shown in Fig. 4(b). Furthermore, Fig. 5 shows the streamwise velocity profile, normal to the surface at X/h = 5, where X denotes the axial distance and $h$ represents the step height. The results obtained in the present study match well with existing DSMC results obtained by Guo et al [16].

**Table 6 Simulation parameters of Guo et al. [16]**

| H | $P_\infty$ | $T_\infty$ | $Ma_\infty$ | Step height | Mean free path ($\lambda$) | Collision frequency | Inflow density |
|---|---|---|---|---|---|---|---|
| (Km) | (Pa) | (K) | - | (mm) | (m) | (1/s) | (kg/m$^3$) |
| 30 | 1197 | 226.5 | 6 | 10 | $4.41 \times 10^{-6}$ | $9.22 \times 10^7$ | $1.520 \times 10^{-5}$ |

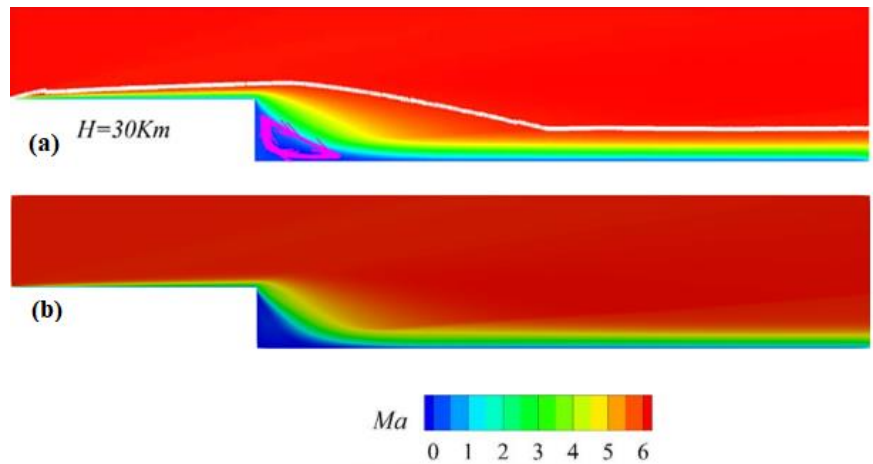

**Fig. 4 Comparison of velocity contour for the case of H=30 km (a) Guo et al. [16] (b) present study.**

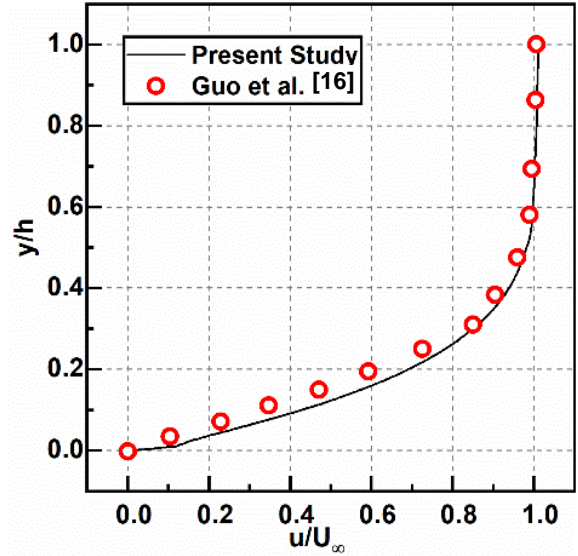

**Fig. 5** Streamwise velocity along the vertical line of $X/h = 5$ for the case of H=30 km.

## 4.2 Chemically reacting flow over a circular cylinder

The two-dimensional hypersonic rarefied airflow over a circular cylinder with a diameter of 2 m, investigated by Scanlon et al. [36], is a benchmark case for the Q–K chemical reaction model. The free-stream and flow conditions for atmospheric altitude, H = 86Km, are shown in Table 7. Fig. 6(a) compares Mach number contours reported in [36] and the present simulation. Fig. 6(b) depicts the species' number density along the stagnation streamline in front of the cylinder. The results obtained match well with the Q-K results obtained in [36], thus, supporting the validity of our results for chemically reacting rarefied flow.

**Table 7 Simulation parameters of Scanlon et al. [36]**

| H | $T_\infty$ | $Ma_\infty$ | Cylinder Diameter | Mean free path | Collision frequency | Inflow density |
|---|---|---|---|---|---|---|
| (Km) | (K) | - | (m) | (m) | (1/s) | (kg/m³) |
| 86 | 187 | 25 | 2 | $1.23 \times 10^{-2}$ | $2.96 \times 10^4$ | $1.43 \times 10^{-5}$ |

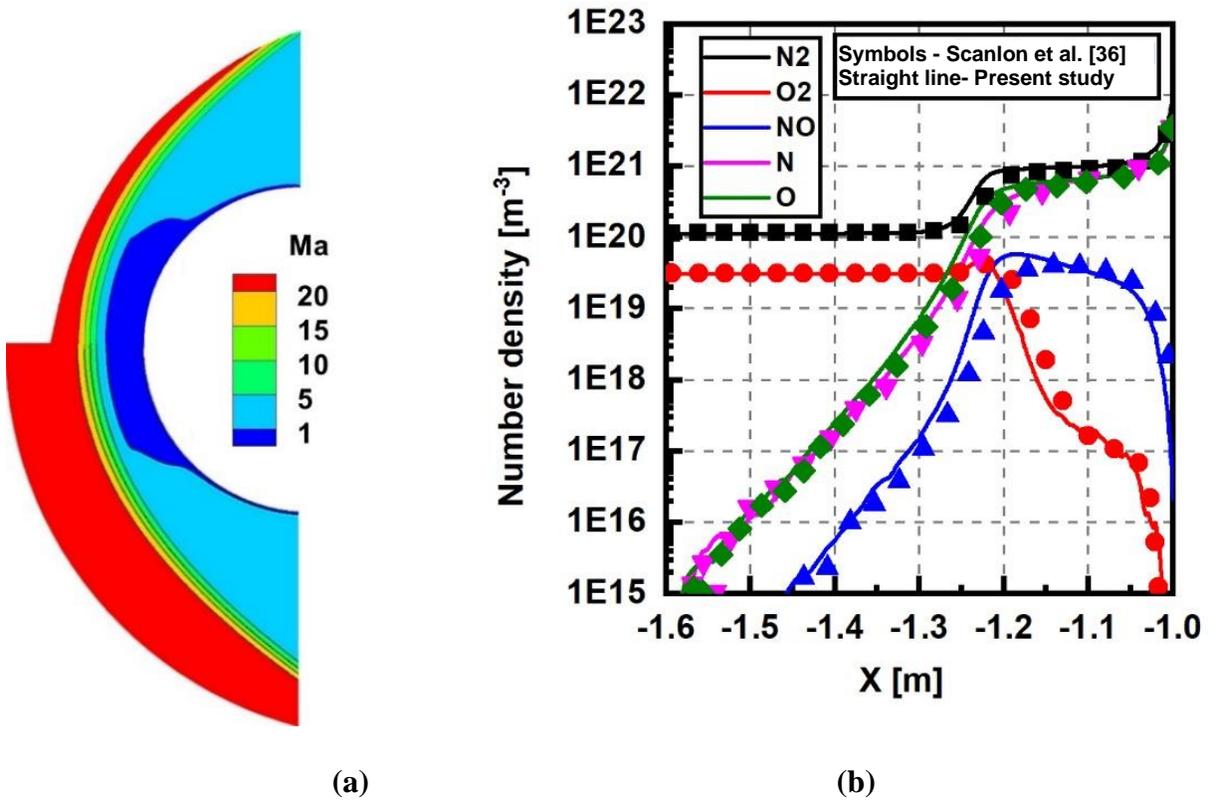

(a)                                                   (b)

**Fig. 6   (a) Comparison of overall Mach number contour between published Q-K results by Scanlon et al. [36] (lower half) and present simulation results (upper half), (b) Comparison of number density along the stagnation streamline.**

# Chapter 5

# Results and Discussion

To study the influence of rarefaction, five distinct Knudsen numbers in various rarefaction regimes were simulated, and the flow properties were analysed. For all the instances, the free-stream Mach number considered was 25.

## 5.1 Flow properties

*5.1.1 Temperature Field*

*Temperature Contours*

The contours of overall temperature, translational temperature, rotational temperature, and vibrational temperature normalized with free-stream temperature for $Kn = 1.06$ are plotted in Fig. 7(a-d). The contours show that the translational temperature near the wall is much greater than the temperature imposed at the wall and is of the order of $10^3 \ K$. This difference can be attributed to the viscous dissipation effects seen in the shear layer at the step height and near the wall. Similar behaviour is observed for rotational and vibrational temperatures as well. The combined effect is conveyed in the overall temperature contour in Fig. 7(a). Just upstream of the step, the flow begins to expand rapidly. This results in expansion cooling in the vicinity of the step and hence lower values of translational and rotational temperature, as can be seen from the cold region behind the step in Fig. 7(b) & Fig. 7(c). Fig. 7(d) shows that the vibrational excitations behind the step remain unaffected and as such, no marked difference is observed in the vibrational temperature. Furthermore, the effects of expansion cooling are also observed downstream near the outlet as a reduction in translational temperature (Fig. 7(b)). As the flow reaches the outlet, the high translational temperature in the shear layer just downstream of the step reduces owing to diffusion, advection, and reduced shear rates. Contrastingly, the rotational and vibrational temperatures are significantly amplified near the outlet owing to the non-equilibrium effects as visible in Fig. 7(c) & Fig. 7(d). The combined effect of this behaviour is reflected in the overall temperature in Fig. 7(a) with a relatively uniform temperature variation downstream of the step, where the decreased translational temperature is compensated by the combined effect of increased rotational and vibrational temperature.

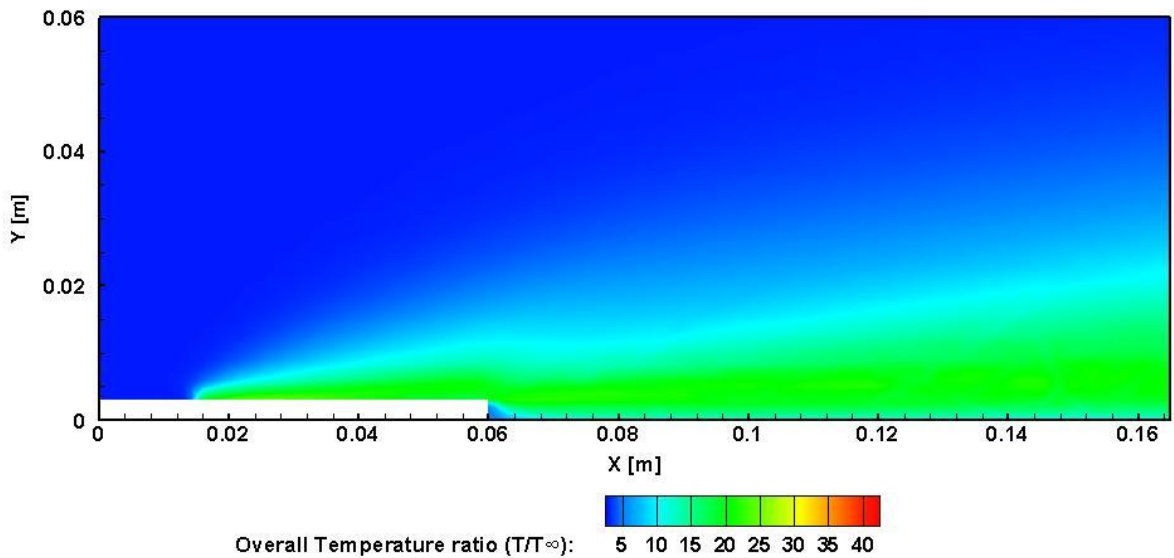

(a)

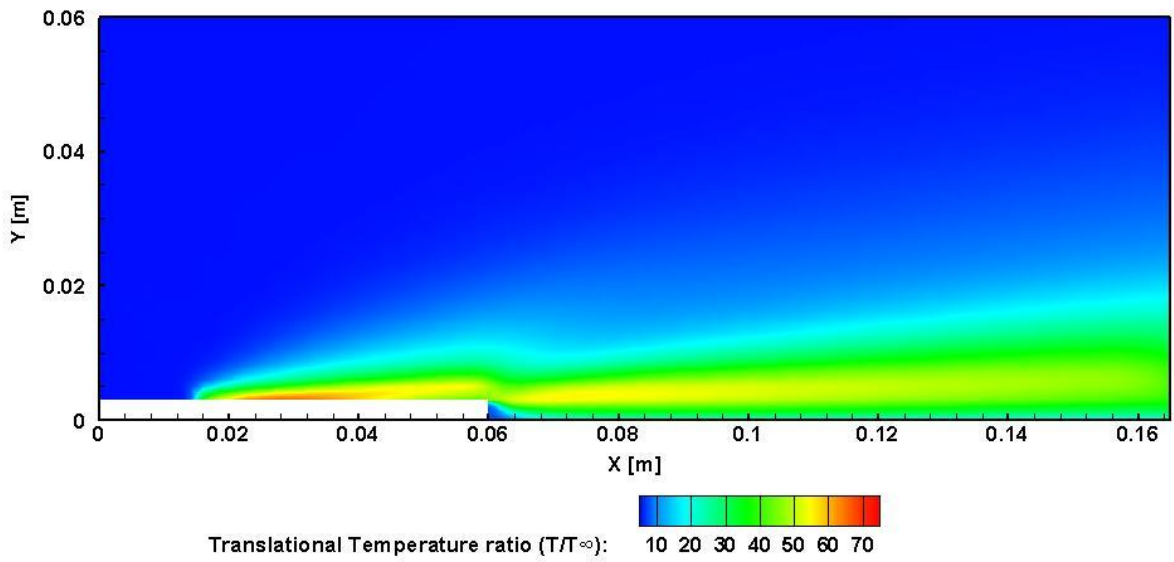

**(b)**

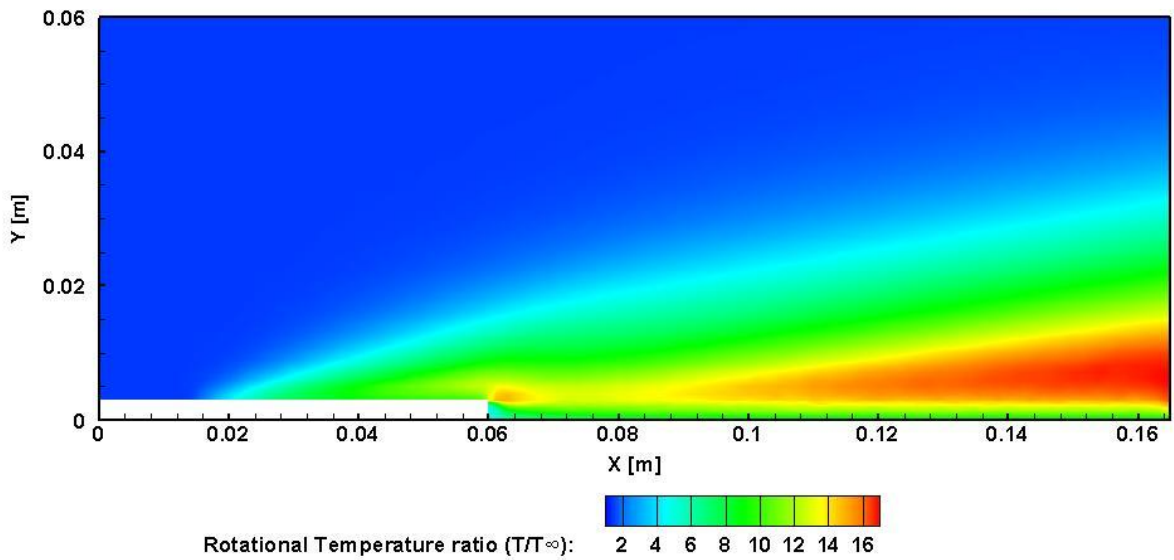

**(c)**

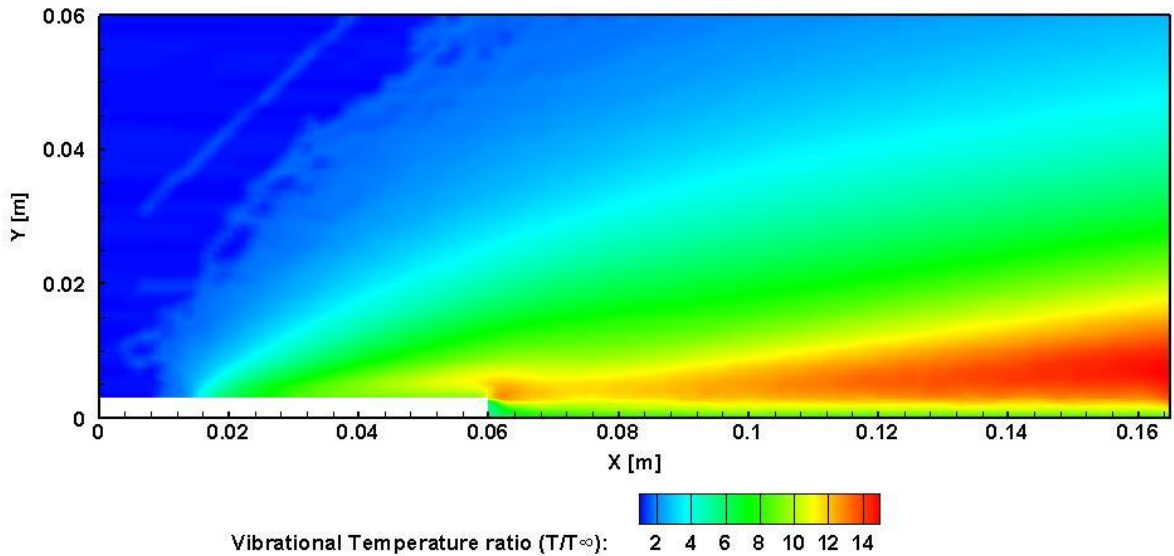

**(d)**

**Fig. 7    Normalised temperature contours of the gas mixture for $Kn = 1$ (a) Overall Temperature, (b) Translational Temperature, (c) Rotational Temperature, and (d) Vibrational Temperature**

*Temperature Profiles*

The locations labelled **a**, **b**, **c**, and **d**, in Fig. 2, represent four different X locations, namely, X=30 mm, 59 mm, 61 mm, and 120 mm, respectively where the temperature profiles are studied. The profiles are also analysed along the horizontal direction for three different segments namely $Y/H = 0, 0.025$ and $0.5$, which are the bottom surface, the centre of the step, and the centre of the flow domain, respectively.

Fig. 8(a-d) compares the overall temperature profiles normalised with the freestream temperature at different $X$ locations. The temperature profiles for both chemically reacting flow as well as non-reacting flow are compared with each other. In all the plots of the chemically reacting

flow, the temperature observed near the wall is an order higher than the imposed wall temperature, and even greater temperature values are observed at the step height in the shear layer due to viscous dissipation. Except for $X = 61\ mm$ which is just downstream of the step, the temperature ratio at the wall increases with rarefaction. At higher $Kn$, viscous heat generation is distributed over a smaller number of molecules, thereby resulting in a greater temperature increment. At the middle of the flow domain, $Y = 30\ mm$, the temperature again rises with rarefaction. The general trend observed along the Y direction, away from the wall, is the rapid increase in temperature until it attains a maximum value, and then follows a gradual decrease in the temperature ratio until the ratio equals 1.

When the results are compared with the non-reacting flow, a considerable difference is observed. At $X = 30\ mm$ the wall temperature ratio values predicted by non-reacting flow, differ with chemically reacting flow by roughly 78%, 109%, 57.2%, 0.49% and -5.7% for $Kn = 0.05, 0.10, 1.06, 10.33$ and $21.10$, respectively. Similarly, just upstream of the step at $X = 59\ mm$, the wall temperatures differ by 50.8%, 67.6%, 71%, 16.7% and 5.2% for $Kn = 0.05, 0.10, 1.06, 10.33$ and $21.10$, respectively with respect to the chemically reacting flow. At $X = 61\ mm$, the percentage error is 5.2%, 11.6%, 6.08%, -4.7%. Finally, at $X = 120\ mm$, the error in the wall temperature ratio is between 18-19% for all degrees of rarefaction. At all $X$ locations, the free molecular flow shows a good agreement with the results predicted by the non-reacting flow. Furthermore, the maximum temperature in the shear layer also is calculated and the error is measured with the non-reacting flow for different locations. At $X = 30\ mm$ the maximum temperature ratio values predicted by non-reacting flow, differ with chemically reacting flow by roughly 72%, 88.9%, 57.2%, 0.49% and -5.7% for $Kn = 0.05, 0.10, 1.06, 10.33$ and $21.10$, respectively. At $X = 59\ mm$, the percentage error is 46.9%, 57.8%, 88.7%, 16.7%, and 5.2% for $Kn = 0.05, 0.10, 1.06, 10.33$ and $21.10$, respectively. Downstream of the step, the error in the maximum temperature is greater than 40% for all $Kn$ at $X = 61\ mm$ and $120\ mm$.

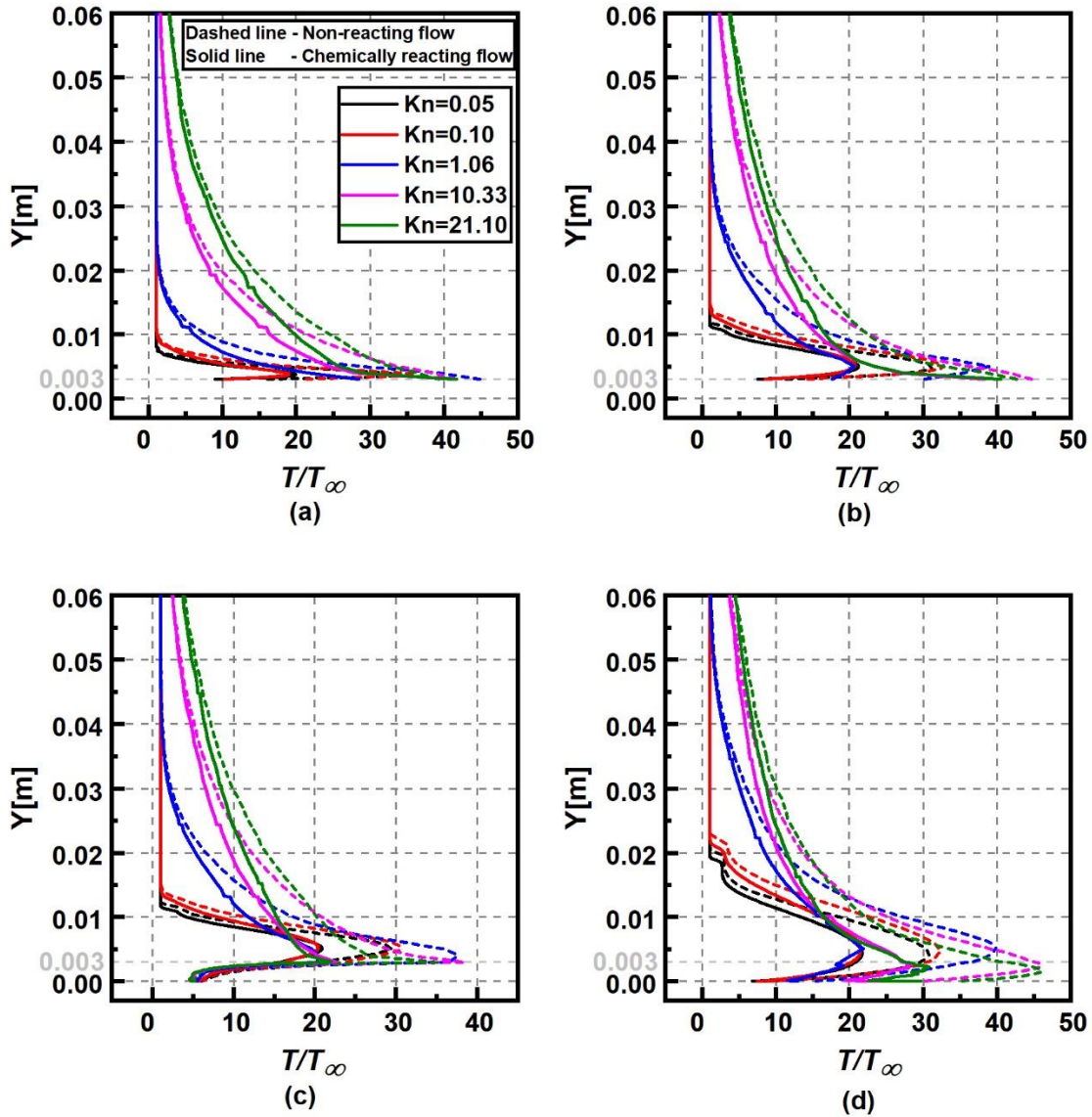

**Fig. 8**　Variation of the non-dimensional overall temperature perpendicular to the surface of BFS for different $Kn$ at (a) X = 30 mm, (b) X = 59 mm, (c) X = 61 mm, and (d) X = 120 mm.

Fig. 9(a-c) compares the temperature ratio distribution along the length of BFS for different $Kn$ at bottom surface ($Y/H = 0$), the centre of the step ($Y/H = 0.025$), and the centre of the flow domain ($Y/H = 0.5$). At the bottom of the step, the temperature ratio increases rapidly along the flow direction due to viscous dissipation effects for $Kn = 0.05, 0.10$ & $1.06$, until it attains a constant value. Whereas, for $Kn = 10.33$ and $21.10$, the temperature ratio rapidly increases for the same reason, until an inflection point, from where it gradually increases again. At any location downstream of the step, the temperature ratio increases with rarefaction. At the centre of the step, $Kn = 10.33$ and $21.10$ cases show a similar trend as the bottom surface, until the inflection point, from where it gradually increases and attains a constant value. The magnitude of temperature ratio increases rapidly and gradually decreases for $Kn = 0.05, 0.10$ & $1.06$ cases. At the centre of the flow domain, the temperature ratio is constant and equal to 1 in the slip regime, whereas other regimes show variation along the flow direction.

At all the $Y$ locations, on an average, the non-reacting flow over-predicts the temperature ratios. At the bottom of the step, the non-reacting flow overpredicts the value of temperature ratio by about 18-20 % for all $Kn$ on an average. At $Y/H = 0.025$, along the flow direction, the shear effects are more prominent and the average overprediction of temperature ratios for a non-reacting flow are approximately 41.5%, 48.2%, 64%, 63.5% and 67.7% for $Kn = 0.05, 0.1, 1.06, 10.33$ and $21.10$, respectively. For $Kn = 21.10$, the maximum percentage difference is as high as 141.5%. Finally, along the centre of the domain, the temperature ratio values in slip flow regime for both non-reacting and chemically reacting flow closely agree with each other, whereas for transition and free-molecular flow the average overprediction of temperature ratio is between 22-24%.

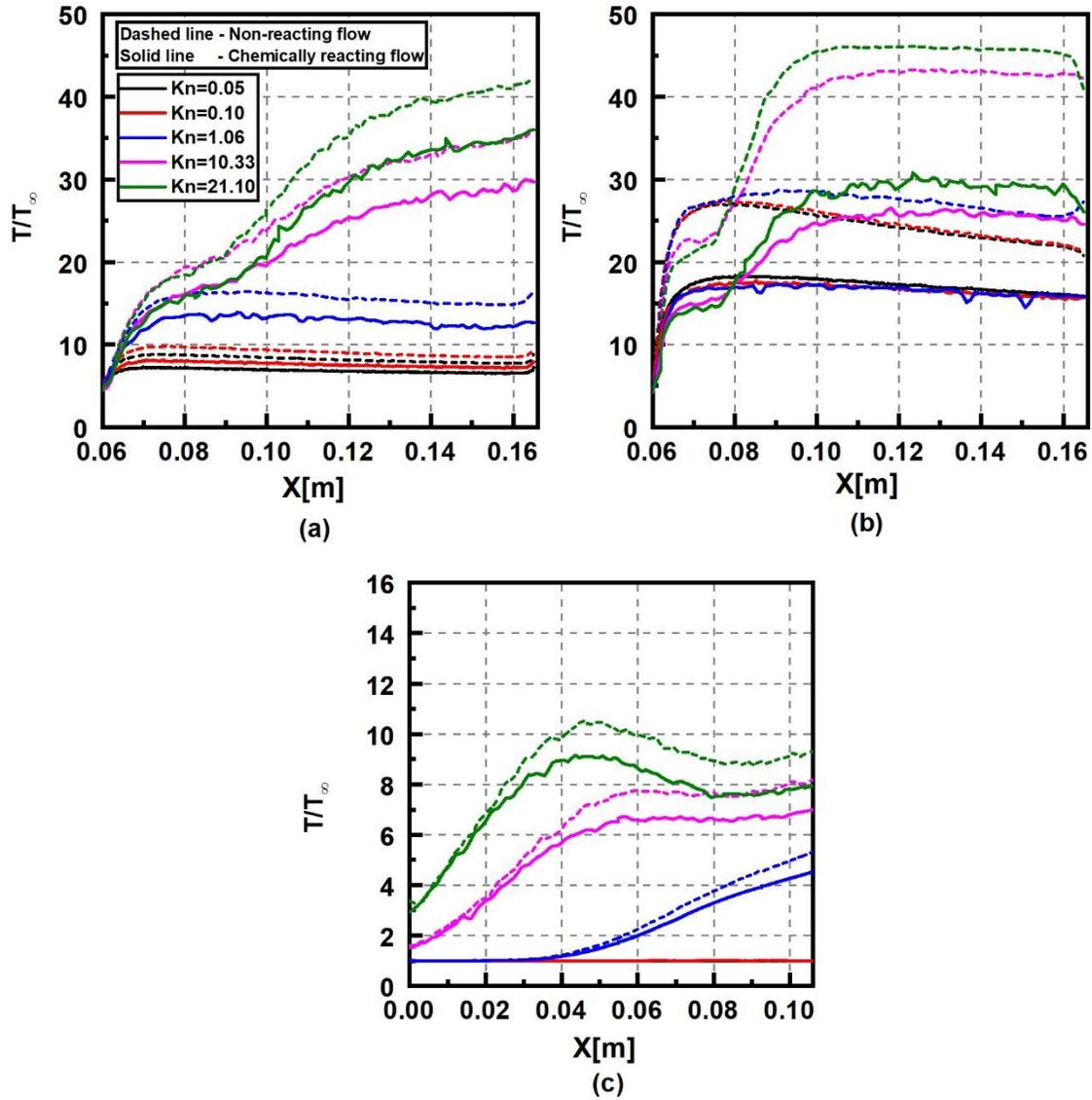

**Fig. 9** Variation of non-dimensional overall temperature along the length of BFS for different $Kn$ at (a) $Y/H = 0$, (b) $Y/H = 0.025$, and (c) $Y/H = 0.5$.

# Chapter 6

# Conclusions

In the present study, the temperature distribution of both chemically reacting, as well as non-reacting hypersonic rarefied gas flow over a BFS is investigated in all the rarefaction regimes using the DSMC method. Q-K model was used to simulate a 5 species air model with 19 reactions. From the present study, the following conclusions can be drawn:

1. The shear effects have a profound effect on the temperature distribution.
2. When chemical reactions are not considered, the mean percentage difference of temperature values between chemically reacting and non-reacting flow, near the downstream wall of BFS is between 18-20% for all rarefaction regimes.
3. Near the shear layer, the average over-prediction of temperature ratios by non-reacting flow is above 40% for all rarefaction regimes.

## 6.1 Future Scope

The results presented above help gain more insight into the development of the thermal boundary layer and quantify the difference in temperature ratios with and without chemical reactions. Future work can be extended to studying the chemical reaction effects in an open cavity, forward-facing step and over a wall-mounted cube.

# List of Publications

**Singh, A.**, Nabapure, D., and Ram Chandra Murthy, K. "Chemically-reacting rarefied hypersonic flow over a backward-facing step." *2021 AIAA Aviation Forum*. – (Extended abstract submitted)

Nabapure, D., **Singh, A.**, Guha, P., and Ram Chandra Murthy, K. "Numerical investigation of rarefied flow over backward-facing step in different rarefaction regimes using Direct Simulation Monte Carlo." *International Journal of Thermal Sciences*. – (in draft)